\documentclass{article}

\usepackage{graphicx}
\usepackage{amsmath,amssymb}

\DeclareMathOperator*\argmax{arg\,max}
\DeclareMathOperator\tr{Tr}
\newcommand\e{\mathrm{e}}
\newcommand\F{\mathit F}
\newcommand\ii{\mathrm{i}}
\newcommand\up{\boldsymbol p}
\newcommand\prob{\mathbb P}
\newcommand\R{\mathrm P}

\newcommand\Sph{\mathcal S}
\newcommand\uk{{\boldsymbol k}}
\newcommand\scha{\Delta}
\newcommand\eref[1]{\eqref{#1}}
\newcommand\Eref[1]{Equation~\eqref{#1}}
\newcommand\rmd{\mathrm{d}}

\title{Stochastic description of geometric phase 
for polarized waves in random media}

\author{\textbf{J{\'e}r{\'e}mie Boulanger, Nicolas Le Bihan} \\ 
\normalsize GIPSA-Lab, Dpt. Images and Signal, 11 Rue des Math\'ematiques,\\
\normalsize Domaine Universitaire, 38402 Saint Martin d'H\`eres CEDEX, France \\[2.5ex] %
\large\textbf{Vincent Rossetto} \\ %
\normalsize Universit\'e Grenoble 1 / CNRS, LPMMC, Maison des Magist\`eres, \\
\normalsize 25 avenue des Martyrs, BP 166, 38042 Grenoble CEDEX 9, France}
\date{\today}

\begin{document}
\maketitle

\begin{abstract}
We present a stochastic description of multiple scattering of polarized waves
in the regime of forward scattering. In this regime, if the source 
is polarized, polarization
survives along a few transport mean free paths, making it possible to
measure an outgoing polarization distribution.
We solve the direct problem using compound Poisson processes on the rotation 
group $SO(3)$ and non-commutative harmonic analysis. 
The obtained solution generalizes previous works in multiple scattering
theory and is used to design an algorithm solving the inverse
problem of estimating the scattering properties of the medium from the
observations. This technique applies to thin disordered layers, 
spatially fluctuating media and multiple scattering systems
and is based on the polarization but not on the signal amplitude.
We suggest that it can be used as a non invasive testing method.
\end{abstract}


\section{Introduction}
Soon after its discovery in adiabatic quantum systems by Berry
\cite{Berry1984}, experimental evidence of a geometric phase for polarized
light travelling along a bended optical fiber was shown by Tomita and Chiao 
\cite{Tomita1986}. It was rapidly realized that any classical transverse
polarized wave could exhibit a geometric phase \cite{Segert1987} when
travelling along a curved path in a three dimensions space. 
This, however, is only possible if the wave remains
polarized, and if the direction of propagation evolves smoothly with time.

If one emits a plane monochromatic wave into a scattering medium,
the multiple scattering events will result in a spreading of the
distribution of the wave vector with time~\cite{vandehulst}. 
Depending on the strength
and density of the scatterers, this spreading will happen at different
rates. The correlation length of the wave vector is called the
{\em transport mean free path}~$\ell^\star$. It is directly related,
in the multiple scattering regime, to the effective diffusion constant
of the energy in the system $D=c\ell^\star/3$ \cite{pine1988} ($c$ is the
velocity of the wave).
Several multiple scattering regimes exist,
depending on the ratio between the wavelength $\lambda$, the size $a$ of the
scatterers, the mean free path~$\ell$, the transport mean free
path~$\ell^\star$ and the depth of the
medium~$L$~\cite{ishimaru1978}. In this article, we will investigate the 
regime
\begin{equation}
 \lambda < a \ll \ell \ll \ell^\star \lesssim L .
\end{equation}
The condition~$\ell\ll\ell^\star$ states that the wave is preferentially
scattered to a direction close to the incoming one. 
This is the so called ``forward scattering'' regime.
For spherical scatterers, this is a consequence of the condition~$\lambda<a$.

The transport mean free path~$\ell^\star$ is the length 
scale of disorder in the medium
while the transport mean free path is the scale along which the
waves behave as homogeneous. 
Measuring the transport mean free path is therefore an important issue
in experiments because it is the resolution scale for most of the
physical quantities measurable using waves. 
Several estimation techniques already exist for the transport mean free
path that can be distinguished from the direction of the measured signal.
The simplest idea is to use the intensity transmission ratio~\cite{garcia1992}
to measure~$\ell^\star$ thanks to the diffusion approximation. 
Diffusion models are also useful in anisotropic media~\cite{johnson2008}
and to describe intensity leaks in direction orthogonal to the
incoming source~\cite{leonetti2011}.
In backscattering setup, the opening angle of the weak localization cone,
in which an intensity enhancement due to interferences is observed, 
is related to the transport
mean free path in the medium~\cite{vanalbada1985}. All these experiments
rely on intensity measurements. 
An amplitude independent method has been proposed to measure the
mean free path~$\ell$ from the phase statistics of a Gaussian field
\cite{anache2009}. 

It is known that in the forward scattering regime 
polarization is parallel transported during the
propagation, which may result in the existence of a geometric phase
\cite{rossetto2001}. 
In the eikonal approximation, each ray possesses 
a geometric phase depending on its geometry. At a given point, the
observed phase is not uniquely defined, but has a distribution
related to the distribution of paths from the source. This
distribution was recently shown to depend on the outgoing direction
of the ray \cite{rossetto2009}.

The distribution of geometric phase from multiple scattering of
a polarized incident wave was adressed ten years ago~\cite{krishna2000} without
any condition on the final direction. The calculation was based on Brownian
motion at the surface of a sphere and followed an approach developed earlier
by Antoine {\it et al.}~\cite{antoine1991}. However multiple scattering is 
rigorously a Brownian motion on a sphere only in the limit of very
weak and dense scatterers. A more general model would be a random walk
with macroscopic steps.

We consider media with a thickness~$L$ of the order of a few $\ell^\star$,
therefore the outgoing wave vectors are not evenly distributed. The joint
distribution of direction and polarization state, related to paths
statistics, contains informations concerning the scattering events
that are not yet randomized. The number of scattering events is of the order
of~$L/\ell$ and in the case where this number 
is not very large ($L/\ell\simeq10$)
the usual approximations are not suited.
For this reason, attention was recently brought to stochastic processes 
for the description of a scattering system where the number of scattering
events is small.

Stochastic models have been introduced fifteen years ago 
in multiple scattering \cite{PhysRevE.52.5621} to solve direct problems.
Recently, Le Bihan and Margerin~\cite{Nico09} showed that a stochastic model
for the direction of propagation  can be used to solve
inverse problems. They computed the mean free path in a medium from
the distribution of outgoing directions.
We extend this model to take polarization into account and estimate the
transport mean free path~$\ell^\star$ from the geometric phase distribution.

\section{Scattering of polarized waves}
\label{Section_Polar}

Without loss of generality, we consider linearly polarized waves. 
The results we present remain valid as long as
polarization is not purely circular.

Polarization of transverse waves lies in the plane
orthogonal to the direction of propagation. We describe linearly
polarized plane waves with two vectors: The direction of propagation $\uk$
and the direction of polarization $\up$, with $\up\perp\uk$. 
We consider only media with isotropic uniform absorption,
the wave amplitude is therefore completely determined for all
paths and does not play any relevant role.
The direction $\uk$ lies on the unit sphere~$\Sph^2 \subset{\mathbf R}^3$. 
The direction of polarization~$\up$ lies in the tangent 
plane $T_\uk\Sph^2$. The frame~$\F$ is fully determined by~$\uk$ and
$\up$ and contains all information concerning the polarization
and direction of propagation. (The third vector of~$\F$ is 
indeed~$\uk\times\up$).

We follow the wave state along a ray using the frame $\F$.
Between scattering events, $\F$ remains constant.
The changes of~$\F$ occur at scattering events. As $\F$ is a frame,
any change, or \emph{jump}, of~$\F$ corresponds to a rotation matrix.
The scattering angle~$\theta$ is random, following a probability distribution
function~(\textit{pdf}) called \emph{phase function}~$\Phi(\theta)$ that
depends on~$a/\lambda$ and on the nature of the scatterer. 
The average value
of~$\left\langle\cos\theta\right\rangle= 
   \int\Phi(\theta)\cos\theta\sin\theta\rmd\theta$ 
is called the \emph{scattering anisotropy} and
is noted~$g$. 
In the forward scattering regime, the
main scattering angle~$\theta$ is small and~$g$ is close to~$1$.
We use the ZYZ convention for the Euler angles
describing the rotations of $SO(3)$ and note the angles 
$(\psi, \theta, \varphi) \in \left]-\pi,\pi\right]\times
\left]0,\pi\right]\times\left]-\pi,\pi\right]$. 

\begin{figure}
\begin{center}
\includegraphics[width=0.75\textwidth]{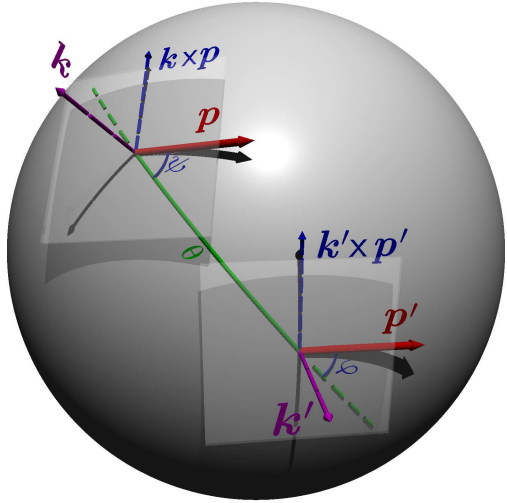}
\caption{Parallel transport of a vector along the geodesic of length
$\theta$ (solid line). The parallel transport constraint results in the
fact that $\psi$ and $\varphi$ are equal (algebrically opposite).
\label{parallel}} 
\end{center}
\end{figure}

The rotation matrix~$\R$ acting on the frame~$\F$ at a scattering
event is fully determined by the incoming
direction~$\uk$ and the outgoing direction~$\uk'$ and the
requirement that the vector~$\up$ is parallel transported~\cite{rossetto2001}. 
On the unit sphere, jumps of the stochastic process 
are represented by geodesics (arcs of great
circles). Moreover, parallel transport in the direct space
implies parallel transport in the phase space from~$T_\uk\Sph^2$ 
to~$T_{\uk'}\Sph^2$ \cite{Jordan2010} and therefore
that the angle between~$\up$ and the geodesic remains constant.
As a consequence the Euler angles $\psi$ and $\varphi$ of the rotation matrix
$\R$ are related by (see figure \ref{parallel})
\begin{equation}
 \psi=-\varphi.
\end{equation}

Parallel transport of polarization through a scattering event may 
therefore be described by a rotation matrix~$\R(\theta,\psi)$,
$\theta$ is the length of the geodesic 
on the unit sphere and~$\psi$ is the angle between this geodesic 
and the polarization vector~$\up$. The relation between the
incoming and outgoing frames is
\begin{equation}
\F'=\F\,\R\left(\theta,\,\psi\right).
\label{eq_frame}
\end{equation}

Note that the rotation matrix~$\R$ acts on the right of~$\F$. 
An expression in terms of {\em left} action of rotation matrices can be
obtained, but would lead to a random process over $SO(3)$ with
{\em dependent} increments. The independence of increments will
be useful for the repeated action~\eref{eq_frame} on~$\F$ that we
present in the next section. 

\section{Multiple scattering process}
\label{Section_CPP}
 
We consider a plane wave with linear polarization, corresponding to
the frame $\F_0$ as described in the previous section.
The scattering events are described as
random rotations acting on the right of~$\F_0$.
The frame~$\F_0$ changes according to~\Eref{eq_frame} 
for each scaterring event, so that it becomes 
the frame $\F_n$ after~$n$ scattering events:
\begin{equation}
\F_n=\F_0\R\left(\theta_1,\,\psi_1\right)
  \R\left(\theta_2,\,\psi_2\right)\cdots
  \R\left(\theta_{n},\,\psi_{n}\right).
\label{frame_F_n}
\end{equation}
The scattering event are independent and the scatterers
are identical, so that the random rotations 
$\R\left(\theta_m,\,\psi_m\right)$ are independent
and identically distributed. 

From the source to the observer, a wave may encounter a variable number of
scatterers. The
number of scattering events in a time $t$ can be taken as a Poisson process
$N(t)$ \cite{Nico09}. The Poisson parameter $\eta$ is directed related to the
mean free path $\ell$ by $\eta=c/\ell$.
All these considerations result in the process~$\F_t$ which we define
as 
\begin{equation}
\F_t=\F_0 \prod_{k=1}^{N(t)}
    \R\left(\theta_k,\,\psi_k\right),
\label{Model_CPP_TP}
\end{equation}
where $\displaystyle{\prod}$ denotes the right-sided product on $SO(3)$. This
equation expresses the stochastic process $\F_t$ as a
compound Poisson process (CPP) on $SO(3)$ with parallel transport.
We denote by~$p_\nu$ the distribution of the process~$\F_t$ at a given
time~$t$, with $\nu=\eta t$.
A Poisson process is a pure jump process with independent increments
and we have
\begin{equation}
p_\nu=\sum_{n=0}^{\infty} \e^{-\nu}\,\frac{\nu^n}{n!} \; p_{\F_n}.
\label{p nu}
\end{equation} 
where $\e^{-\nu} \frac{\nu^n}{n!}$ 
is the probability to have exactly $n$
scattering events in the time interval $[0,t]$.
Thanks to \Eref{frame_F_n} and the 
independence of the parallel transport rotations, 
$p_{\F_n}$ is given by~\cite{Chirikjian2001}:
\begin{equation}
p_{\F_n}= p_{\F_0} \ast p_{\R_1} \ast \cdots 
           \ast p_{\R_n} = \Phi^{\ast[n]}.
\label{convol_R}
\end{equation}
Here the symbol $\ast$ is the convolution product 
on~$SO(3)$ and~$\Phi^{\ast[n]}$ denotes the result of
the convolution of $n$ identical functions~$\Phi$.
We have used the initial condition $p_{\F_0}=\delta\left(I_3-\F_0\right)$. 
In the case where the source is a distribution of directions of
propagation and polarization, the distribution~$p_{\F_n}$ is 
the convolution of the initial distribution~$p_{\F_0}$ with
$\Phi^{\ast[n]}$. 


\section{Harmonic analysis}
We use harmonic analysis on~$SO(3)$ to transform 
the expression~\eref{convol_R}. Harmonic analysis on~$SO(3)$ is
similar to Fourier series and in particular transforms convolutions
into products. 
As in Ref.~\cite{Nico09}, we get a semi-analytic expression 
of $p_\nu$ that we use for statistical estimation in 
Section~\ref{Section_Estimation}.
 
Th Fourier basis of functions~$f\in L^2(SO(3),{\mathbf R})$
is made of the Wigner $D$-matrices $(D^j)_{j\geqslant 0}$
\cite{Dieu80,Chirikjian2001}. The Wigner $D$-matrices are
$(2j+1)\times(2j+1)$ square matrices with coefficients
\begin{equation}
 D^j_{m,n}(\psi,\theta,\,\varphi)
=\e^{-\ii m\psi}d^j_{m,n}(\theta)\e^{-\ii n\varphi}.
\label{defWigner}
\end{equation}
The Fourier coefficients are matrices of the same size
which coefficients are defined as 
$\widetilde{f}^j_{m,n} = 
  \left\langle \Phi \mid D^j_{m,n} \right\rangle$, 
where $\langle\;\rangle$ represents
the scalar product for function on $SO(3)$ \cite{Dieu80}. 
Like in the circular Fourier theory, the transform of a convolution
is a product (here, a matrix product)
\begin{equation}
\widetilde{f_1\ast f_2}^j=\widetilde{f_1}^j \widetilde{f_2}^j. 
\end{equation}
We deduce the Fourier coefficients of~$p_\nu$ from 
\Eref{p nu}
\begin{equation}
\widetilde{p_\nu}^j=\sum_{n=0}^{\infty}\e^{-\nu} \frac{\nu^n}{n!}
  \left(\widetilde\Phi^j\right)^n 
  =\exp\left[\nu(\widetilde\Phi^j-I_{2j+1})\right],
\label{coeff_pFT}
\end{equation}
where $\exp$ denotes the matrix exponential. 
Using the notation $\scha(x)=2\pi\sum_k\delta(x-2\pi k)$
to put the parallel transport constraint on the Fourier coefficients
we obtain
\begin{equation}
 \widetilde{\Phi}^j_{m,n}=\frac{1}{8\pi^2}
\iiint 
     \Phi(\theta) \scha(\varphi+\psi) D^j_{m,n}(\psi,\,\theta,\,\varphi)
     \sin\theta\rmd\theta\,\rmd\varphi\,\rmd\psi.
\end{equation}
We can simplify using the definition~\eref{defWigner} 
\begin{equation}
\widetilde{\Phi}^j_{m,n}=\frac{\delta_{mn}}{2}
  \int_0^\pi \Phi(\theta)d^j_{m,m}(\theta)\sin\theta\rmd\theta,
\end{equation}
and show that the Fourier matrices $\widetilde{\Phi}^j$ are diagonal
for all $j$.  Thanks to
\Eref{coeff_pFT}, it is clear that $\widetilde{p_\nu}^j$ is
also diagonal at all orders~$j$.

Using the {\em inverse} Fourier formula \cite{Dieu80}, $p_\nu$ reads:
\begin{equation}
p_\nu=\frac{1}{2\pi}\sum_{j=0}^{\infty} 
  (2j+1)\tr(\widetilde{p_\nu}^j D^{j \dagger})
\end{equation}
where $\tr$ denotes the trace and the ${}^{\dagger}$ denotes the Hermitian 
conjugation of elements from $\mathcal{M}_{2j+1}(\mathbf{C})$.

The variable $\beta=\psi-\varphi$
represents the direction of polarization of the wave after its propagation
through the medium, which is the {\em geometric phase}. Thanks to symmetry,
the distribution of~$\beta$ is the same as the one of~$\psi+\varphi$.
Combined with the fact that 
$\widetilde{p_\nu}^j$ is diagonal, 
we conclude that~$p_\nu$ is a function of~$\theta$ and $\beta$
\begin{gather}
p_\nu(\theta,\,\beta)=R_0(\theta,\,\nu) 
     + 2 \sum_{m\geq 1} \cos(m\beta)
         R_m(\theta,\,\nu),
\label{proba phase}\\
R_m(\theta,\,\nu)=\frac1{2\pi}
    \sum_{j \geq m} (2j+1) \e^{\nu(\widetilde{\Phi}^j_{m,m}-1)}
    d^j_{m,m}(\theta).
\label{R_m}
\end{gather}

Thus, \Eref{proba phase} is a general expression of the
distribution of geometric phase in multiple scattering regime
for polarized waves.
We have only assumed that the scatterers are spherical and identical.

The limit of Brownian motion, which is made of
isotropic infinitesimal steps of spherical length~$\delta$
occuring at a large rate~$\eta$, corresponds to the physical
situation where scatterers are very weak but have a large spatial
density. For a single step, we get
$\tilde\Phi^j_{m,m}=\exp[-\frac12j(j+1)\delta^2]$. Taking the limit
$\delta\to0$ and~$\eta\to\infty$ such that $\delta^2\nu=\delta^2\eta t=Dt$ 
remains constant, formula~\eref{proba phase} gives the distribution
of the geometric phase for a wave propagating in a heterogeneous
continuous medium. In formula~\eref{proba phase}, one should replace
$\nu(\tilde\Phi^j_{m,m}-1)$ by $-\frac12j(j+1)Dt$. 
This formula was demonstrated by Perrin in his study of rotational Brownian
motion~\cite{perrin1928,perrin1936}, 
while computing the distribution of~$\varphi+\psi$. 

\begin{figure}
\begin{center}
\includegraphics[width=0.8\textwidth]{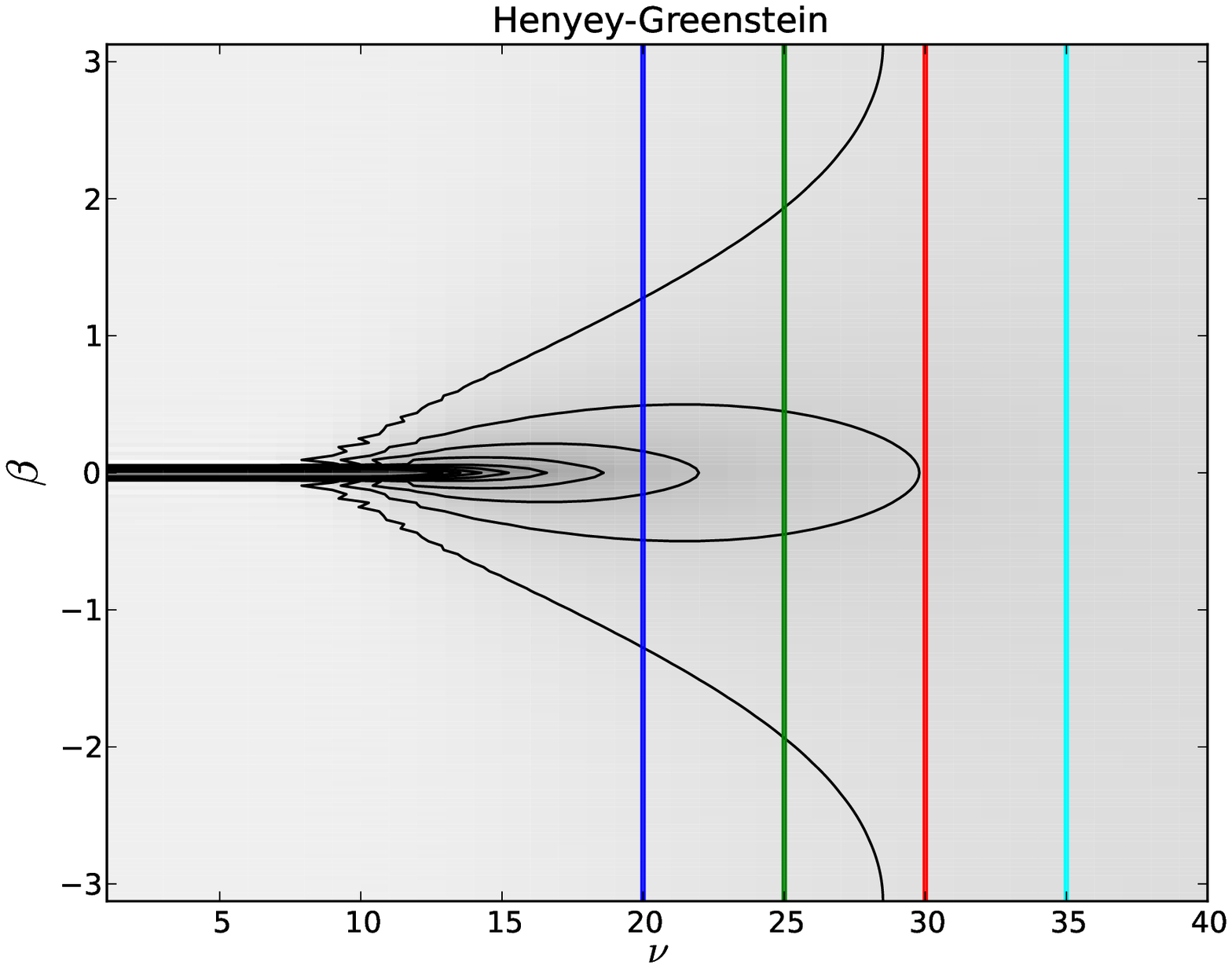} \caption{Evolution
of the density $(p_{\eta t}(\theta=0,\,\beta)$ with time in a medium
containing Henyey-Greenstein scatterers with
anisotropy~$g=0.8$. The curves are iso-density levels from
$0.05$ to $0.45$ by increments of~$0.05$.
The vertical lines correspond to the slices
displayed on Figure~\ref{HG2}. \label{HG1}}
\includegraphics[width=0.8\textwidth]{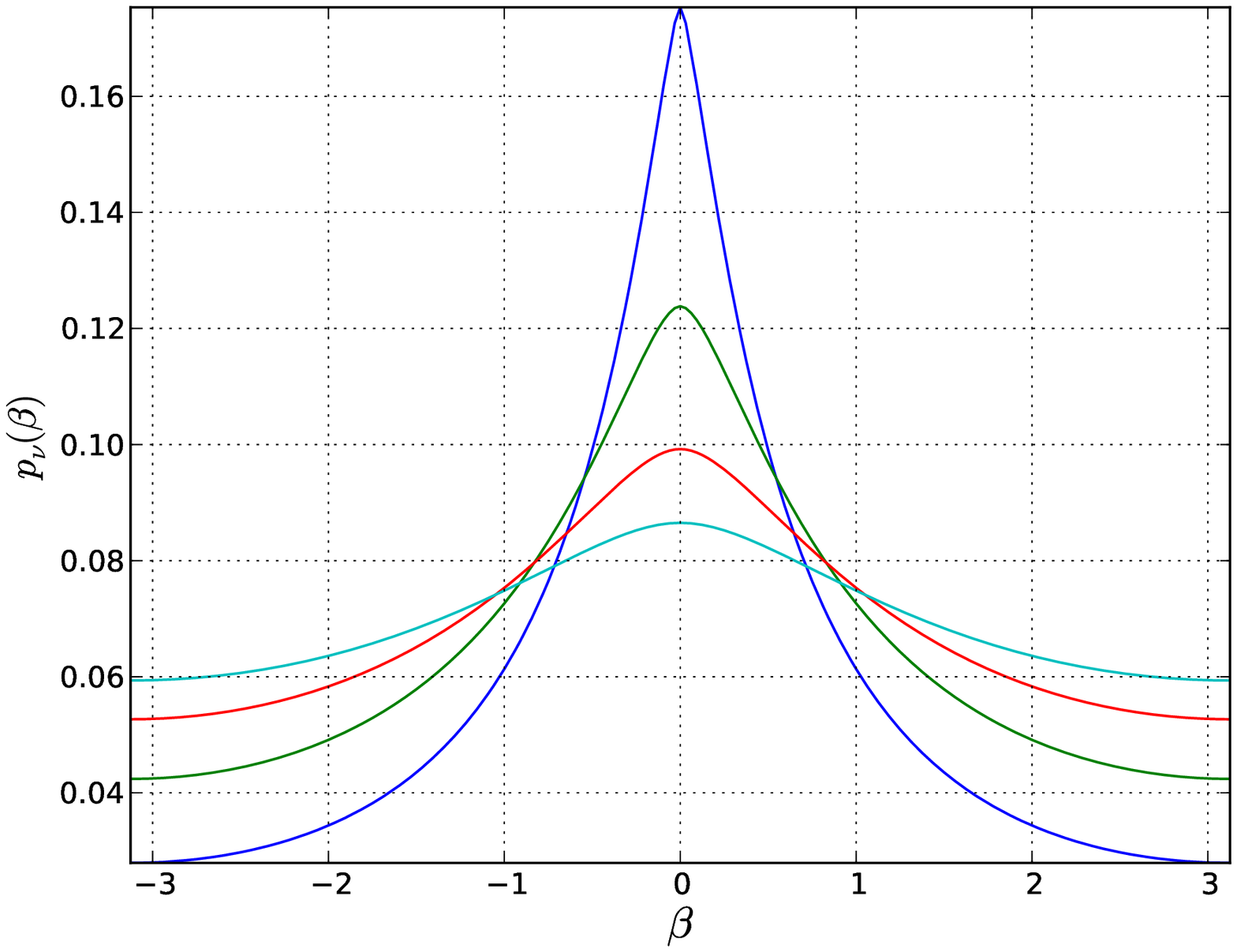} \caption{Distribution
of the geometric phase $\beta$ for the values of the Poisson
parameter~$\nu=\eta t$ : $\nu=20$, $\nu=25$, $\nu=30$ and $\nu=35$. 
corresponding to the vertical lines of Figure~\ref{HG1}.
\label{HG2}}
\end{center}
\end{figure}

\begin{figure}
\begin{center}
\includegraphics[width=0.8\textwidth]{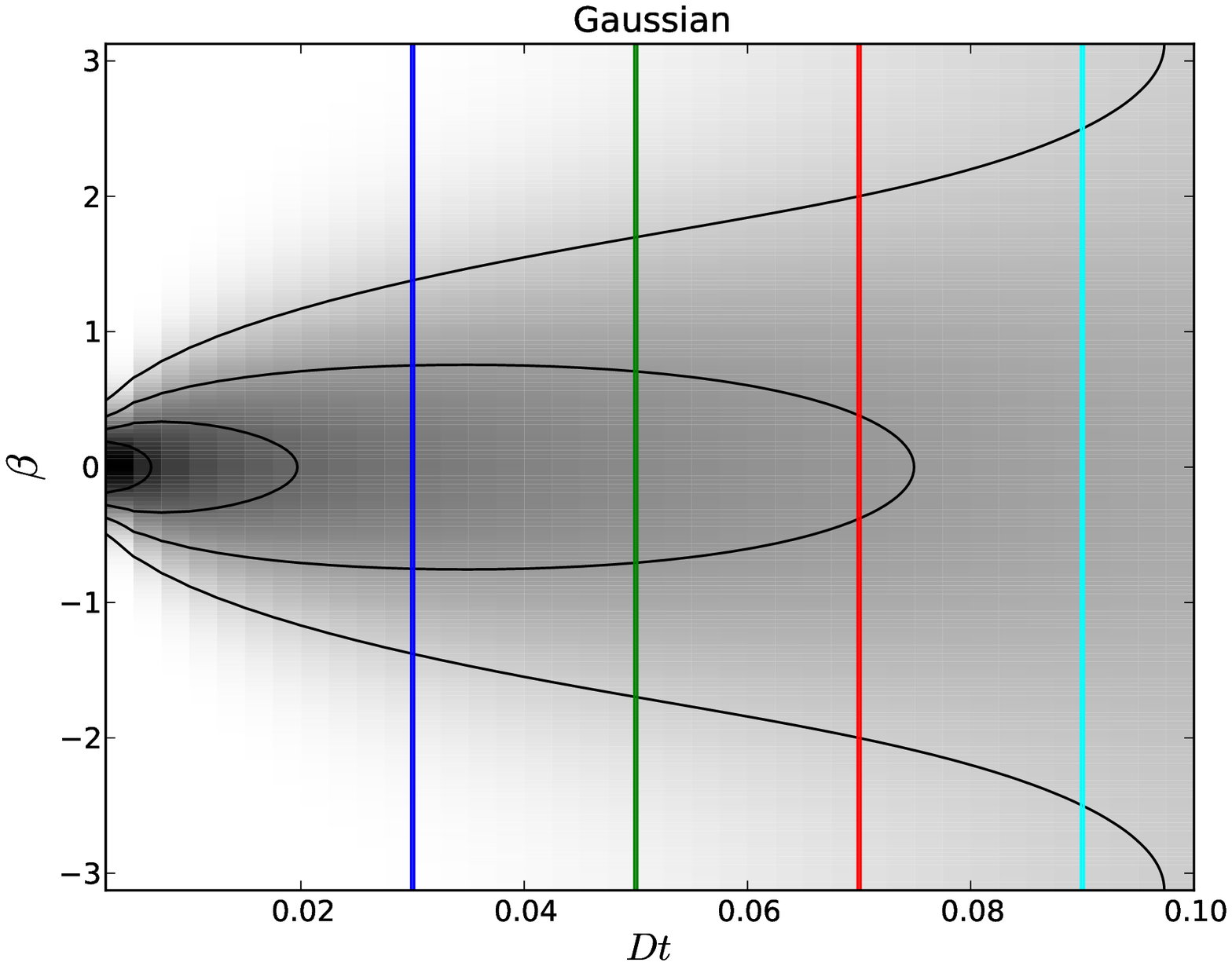}
\caption{Evolution of the density $p_{Dt}(\theta=0,\,\beta)$ with time
for a Gaussian phase function. This corresponds to the
rotational Brownian motion on the sphere studied by Perrin
in Ref.~\cite{perrin1936}. 
The curves are iso-density levels from
$0.05$ to $0.45$ by increments of~$0.05$.
The vertical lines correspond to the slices
displayed on Figure~\ref{G2}. 
\label{G1}}
\includegraphics[width=0.8\textwidth]{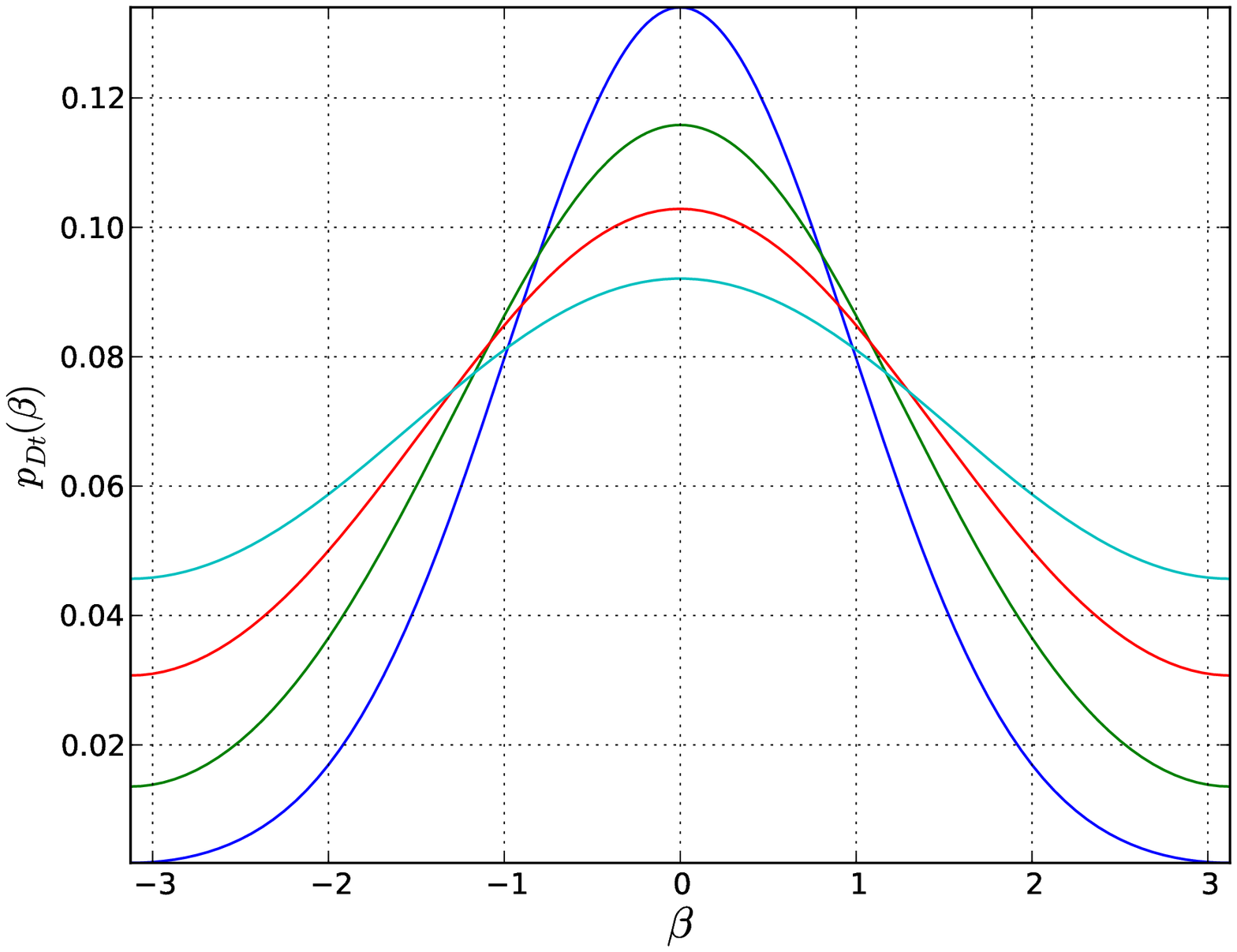}
\caption{Density $p_{Dt}(\theta=0,\,\beta)$ 
for a Gaussian phase function with the values~$Dt=0.03$, $Dt=0.05$,
$Dt=0.07$ and $Dt=0.09$ corresponding to the vertical lines
of Figure~\ref{G1}. 
\label{G2}}
\end{center}
\end{figure}

Taking~$\theta=0$ in~\Eref{proba phase} we get the result demonstrated by 
Antoine~{\it et al.}~\cite{antoine1991},
and integrating over~$\theta$ we find
the result obtained by Krishna~{\it et al.}~\cite{krishna2000}
up to a small difference, ($j(j+1)$ has to be replaced by $j(j+1)-m^2$).
\Eref{proba phase}
is therefore a generalization of these known formulas
that can be applied to an arbitrary phase function, also in the case of a small
number of scattering events and for an arbitrary deviation angle~$\theta$.
The origin of the small difference 
comes from the fact that the Brownian motion of the tangent vector of a
particle moving in space~$\mathbf{R}^3$
has one more degree of freedom than the Brownian motion on~$\Sph^2$ 
(the rotation degree of freedom about the tangent vector). 
In References~\cite{antoine1991,krishna2000} the extra degree
of freedom disappears because these authors considered rotations
with~$\psi=0$ and in this work, we considered rotations with~$\varphi+\psi=0$.
Let us remark that the set of these latter rotations forms a group 
but that the set of rotations with~$\psi=0$ does not.

We display the distribution obtained from formula~\eref{proba phase}
in the two following cases:~discrete Henyey-Greenstein scatterers 
with anisotropy $g=0.8$ on a range of Poisson parameter up to~$\nu=20$ 
on the one hand (Figures \ref{HG1} and \ref{HG2}) and diffusive limit with
$D=1\;\mathrm{rad^2\,s^{-1}}$ on the other hand (Figures \ref{G1} and
\ref{G2}). The exit angle is $\theta=0$ for all of the figures.
We can observe that the behavior of the distributions on Figures
\ref{HG2} and \ref{G2} are very different, especially at short times,
the distribution~$p_\nu$ is sharply peeked for the Henyey-Greenstein
scatterers but in the case of Brownian motion it is smooth. 
At long times, depolarization is observed in both cases. The depolarized
state is reached with significantly distinct behaviours. This is an
illustration that depolarization is significantly dependent on the
scattering properties of the medium. It is therefore natural to
investigate how these scattering properties can be estimated from
observations through signal analysis.

We have shown that the polarization distribution depends on the angle between
the outgoing wave vector~$\uk_{\text{out}}$ and the initial one~$\uk_0$.
Moreover, it actually depends on the position of the last scattering event. Let
us call~$\chi$ the angle coordinate of this point in the cylindrical 
coordinates of axis~$\uk_0$ in the
frame~$\F_0$, and~$(\theta,\,\phi)$ the spherical coordinates
of~$\uk_{\text{out}}$ in the same frame. The distribution of polarization is
then the one given by formula~\eref{proba phase} shifted by an
angle~$(1-\cos\theta)\chi$. 
This shift is due to the fact that the
reference~$\beta=0$ is taken in the parallel transported frame from the source
to the last scattering event, so that an extra phase appears in the
frame~$\F_0$. This extra phase has been already observed in backscattering
configuration ($\theta=\pi$)~\cite{hielscher1997,rossetto2002}.

We remark that in the case where~$\tilde\Phi^j_{m,m}$
depends only on~$j$ (and is noted~$\tilde\Phi^j$),
formula~\eref{proba phase} simplifies to
\begin{equation}
p_\nu(\theta,\beta)=\sum_{j\geqslant0}(2j+1)\e^{\nu(\tilde\Phi^j-1)}
\frac{\sin\left(j+\frac12\right)\omega}{\sin\frac12\omega},
\label{cas symetrique}
\end{equation}
with~$\cos\frac\omega2=\cos\frac\theta2\cos\frac\beta2$. 
As a consequence, there is an unexpected symmetry for~$p_\nu$ 
between the outgoing angle and the geometric phase.

\section{Inverse problem}
\label{Section_Estimation}
In this section, we take advantage of the Fourier expansion and propose a
statistical estimation of the mean free path~$\ell$ obtained from
the measured distribution of the geometric phase $\beta$.
This is achieved by estimating the Poisson parameter~$\nu=\eta t$. 
The estimate of $\nu$ is denoted by $\hat\nu$.

The inverse problem is solved using an expectation-minimization (EM) approach
\cite{Dempster1977} with the parametric CPP.
The EM algorithm is based on the
maximization of the log-likelihood of the {\it a posteriori} distribution given
in \Eref{p nu}. Suppose that we are given a sample of size
$M$ of observations (measurements) $\F_m$, 
$1\leqslant m \leqslant M$. 
We assume that each observation $\F_m$ follows
independently the probability law~\eref{proba phase}, the
joint probability distribution function is
simply the product $\prod_m p_\nu(\F_m)$, with 
$p_\nu(\F)=p_\nu(\theta,\psi+\varphi)$,
where~$\varphi$, $\theta$ and~$\psi$ are the Euler angles of~$\F$.
The value of $\nu$ that maximizes this log-likelihood is
\begin{equation}
\hat\nu = \argmax_\nu \left(\sum_{m=1}^M \log p_\nu(\F_m)\right).
\end{equation}

The EM algorithm is an iterative procedure that almost
surely converges to a local minimum \cite{Dempster1977}. 
The {\em minimization} step (M-step) consists in updating the
estimate $\hat\nu$, while the {\em expectation} step (E-step) consists in
updating $\prob\left(n \mid \F,\,\hat\nu_i\right)
        =p_{\F_n}(\F)/p_{\hat\nu_i}(\F) \times \e^{-\hat\nu_i}\hat\nu_i^n/n!$.
We get a sequence of estimates with the relation
\begin{equation} 
  \hat\nu_{i+1}=\frac{1}{M}
  \sum_{m=1}^{M} \frac 1{p_{\hat\nu_i}(\F_m)} \sum_{n=0}^{\infty} n\, 
    p_{\F_n}(\F_m) \,\e^{-\hat\nu_i}\,\frac{\hat\nu_i^n}{n!}.
\label{sequence nu}
\end{equation}
In practice the sums over $n$ in~\Eref{sequence nu} and 
\Eref{p nu} and over~$j$ in \Eref{R_m}
are truncated to an arbitrarily fixed value~$N$.
At most~$N$ scattering events are described by the model, therefore
$N$ should be taken large compared to~$L/\ell$. 
This EM-algorithm is able to estimate one parameter, the
Poisson parameter~$\nu$ (or equivalently the mean free path
$\ell=ct/\nu$), which means that we have considered that the
phase function~$\Phi$ is known. It is nonetheless
possible to estimate simultaneously any finite number of
parameters of $p_\nu$, such as the anisotropy~$g$ of the scatterers,
but the iteration formulas are more sophisticated.

\begin{figure}
\begin{center}
\includegraphics[width=0.75\textwidth]{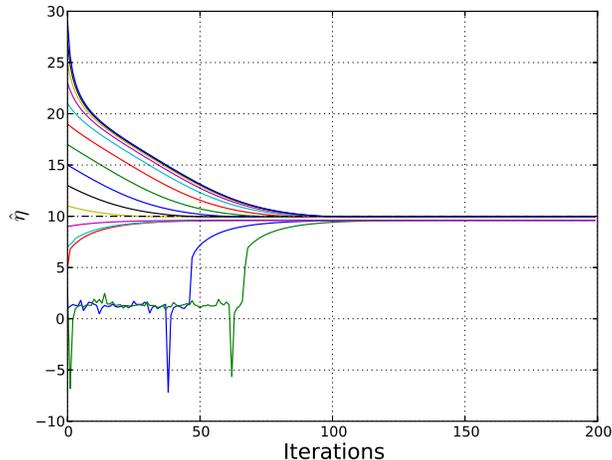}
\caption{Convergence of the estimate of the Poisson parameter
$\hat{\eta}_i$ using EM algorithm with different initialization values 
ranging from $1$ to $29$. The actual value of $\eta$ was set to $10$.
\label{Conv_EM}}
\end{center}
\end{figure}

We illustrate the behaviour of the EM algorithm in Figure~\ref{Conv_EM}
which shows the convergence of the estimator $\hat\nu$ for different
initial values~$\hat\nu_0$. 
Acceleration procedures could be developped in the case
of large dataset from experimental setup. Note that no local minima were
reached in the simulation, leading to convergence in all the
cases. However, as can be seen for initial values inferior to~$10$ in Figure
\ref{Conv_EM}, underestimated initial values lead to a systematic bias in
$\hat\nu$. This suggests that high values should be privileged when
processing dataset, as it does not penalize convergence and leads to a more
accurate estimate. 

\section{Conclusions}
The description of the depolarization of multiply scattered waves can be made,
in the forward scattering regime, through a compound Poisson process (CPP).
This stochastic model predicts the distribution of the geometric phase in
all directions, for discrete or continuous scattering media.
generalizing existing results (forward outgoing direction or
a spatially fluctuating medium). Moreover, the CPP model
allows a more detailed description of the phenomenon as it provides the
behaviour of this distribution as a function of the output scattering angle
$\theta$. The present approach allows to design an iterative
procedure to estimate properties of the scattering medium
through the measurement of the outgoing polarization distribution. We have
illustrated this point by presenting an expectation-minimization (EM) 
algorithm for the estimation of the Poisson parameter, which is directly linked
to the transport mean free path $\ell^\star$. An interesting feature
of this technique is that is relies on polarization rather
than on amplitude measurements. Future work will
consist in validating the proposed model and estimation algorithm on
measurement of polarization distribution for real experiment.


\section*{Acknoledgments}
This work was founded by the CNRS/PEPS grant PANH.

\section*{References}
\bibliography{CPP_SO3_Berry}

\end{document}